\documentclass[reqno,10pt]{amsart}

\newcommand{\metric}{\mathrm{g}}

\begin{document}

\title[Lorentz symmetry breaking]
{Causality, shocks and instabilities in vector field models of Lorentz symmetry breaking.}%
\author{M. A. Clayton}%
\address{Physics Department, Acadia University, Wolfville, Nova Scotia B$0$P $1$X$0$ Canada.}%
\email{michael.clayton@acadiau.ca}%


\date{\today}

\begin{abstract}
We show that that vector field-based models of the ether
generically do not have a Hamiltonian that is bounded from below
in a flat spacetime.  
We also demonstrate that these models possess multiple light cones in flat or curved spacetime, and that the non-lightlike characteristic is associated with an ether degree of freedom that will tend to form shocks.
Since the field equations (and propagation speed) of this mode is singular when the timelike component of the ether vector field vanishes, we demonstrate that linearized analyses about such configurations cannot be trusted to produce robust approximations to the theory. 
\end{abstract}
\maketitle

\section{Introduction}

Recently there has been interest in models in which constants of nature are elevated to the status of fields~\cite{Albrecht+Maguiejo:1999,Barrow:1999,Clayton+Moffat:2001}.
This is partly due to recent experimental evidence for a variation
of the fine structure constant with
time~\cite{Webb+:1998,Webb+:2000,Avelino+:2000,Avelino+Martins+Rocha:2000},
and partly from theoretical interest in, for example, solving the
horizon and flatness problems in
cosmology~\cite{Moffat:1993a,Albrecht+Maguiejo:1999} or modeling
Planck scale deviations from flat spacetime field theory in the very early universe~\cite{Martin+Brandenberger:2000}.

Along with these models, either explicitly or implicitly, comes a
breakdown of Lorentz invariance.  
This appears as contributions to the action that are explicitly written in a
preferred frame~\cite{Maguiejo:2001}, or in a more subtle manner
where, although no preferred frame is chosen, one or more of the
equivalence principles are
violated~\cite{Clayton+Moffat:1999,Clayton+Moffat:1999a,Clayton+Moffat:2001}.
Indeed, the motivation for these models is to embody the idea of a
``varying speed of light'', which should conflict with Einstein's
postulate of the invariance of the propagation
velocity of light.

Among the latter class of models we would include those that have
similar prior-geometric structure~\cite{Drummond:2000}, the
so-called ``$k$-inflation'' and ``$k$-essence''
models~\cite{Armendariz+:1999,Armendariz+:2000},
and vector field models of the
ether~\cite{Jacobson+Mattingly:2000,Jacobson+Mattingly:2000a,Kostelecky+Samuel:1989}.
Although these models are not necessarily intended introduce fields with
dynamically determined light velocities, this nevertheless results from their construction. 
In the same way, higher-dimensional models~\cite{Kiritis:1999,Alexander:1999,Csaki+Erlich+Grojean:2000} also generically end up with multiple light cones as a result of dimensional reduction.

The class of model that we want to discuss here are those
considered, for example,
in~\cite{Jacobson+Mattingly:2000,Jacobson+Mattingly:2000a,Kostelecky+Samuel:1989}.
These models use a (co-)vector field $A_\mu$ to model the
preferred frame representing the ether, introducing a constraint
in the action via a Lagrange multiplier $\lambda$ that forces $A_\mu$ to be
a timelike, null or spacelike vector.   
The model is described by the action:
\begin{equation}\label{eq:genaction}
S= -\int d\mu R
 +\int d\mu\Bigl(-\frac{1}{4}F^2
 +\frac{1}{2}\lambda(A^2-l)
 +A_\mu J^\mu\Bigr),
\end{equation}
where we are using a metric with signature $(+,-,-,-)$, have
chosen $16\pi G=1=c$, and have normalized the vector field $A_\mu$
to remove a possible arbitrary coefficient multiplying the $F^2$
term. We have also written $d\mu=\sqrt{-\metric}d^4x$,
$A^2=\metric^{\mu\nu}A_\mu A_\nu$, and
$F^2=\metric^{\mu\alpha}\metric^{\nu\beta}F_{\mu\nu}F_{\alpha\beta}$.
The source-coupling term $A_\mu J^\mu$ is included to
represent the possible non-derivative coupling of the ether to matter fields
through a matter action: $S_{\mathrm{matter}}(\metric,A,\ldots)$,
and so one may think of $J^\mu=\partial
S_{\mathrm{matter}}/\partial A_\mu$.  

There is a fairly long history of difficulties
associated with coupling vector fields to gravity.
Kucha\v{r}~\cite{Kuchar:1977} has noted that in general, fields that are
derivatively-coupled to gravity may propagate off the light cone
as determined by the spacetime metric.
Examples of this happening when `higher spin' fields were coupled to gravity have been in existence for quite some time~\cite{Buchdahl:1962,Cohen:1967,Velo+Zwanzinger:1969,Aragone+Deser:1971}, and linear models coupling a vector field to gravity are known to possess instabilities unless they are one of Maxwell, Proca, or a purely longitudinal vector field, minimally-coupled to gravity~\cite{Isenberg+Nester:1977}.
In this paper we demonstrate that the models described by~\eqref{eq:genaction} suffer from similar problems, not so much to point out flaws in the literature as to further illustrate the consequences of modifying the structure of vector field models.
As these issues are serious, they play a complimentary role to observational constraints arising from, for example, performing a PPN expansion (see for example~\cite{Will+Nordtvedt:1972a,Will+Nordtvedt:1972b}). 

In Section~\ref{sect:flat}, we begin an examination of the model on a flat spacetime, and show that the constraint imposed by $\lambda$ re-introduces the spatially longitudinal mode of the vector field as a physical field, giving it a propagation speed that depends nonlinearly on the vector field itself.  
We explicitly determine this velocity, and show that for spacelike ether models it is infinite near configurations with $A_0=0$.
In Section~\ref{sect:linearization} we show that the linearization of these models does not correctly reproduce this result, and in Section~\ref{sect:shocks}, we demonstrate that the field equations are also singular at this point.
Also in this section, we show that the the field equations for timelike or spacelike ether models, will tend to form shocks.

We also show,~in Section~\ref{sect:Hamiltonian}, that these models generically do not have a classical energy that is bounded from below regardless of the signature of the ether, and in Section~\ref{sect:others} we briefly consider the implications of adding additional terms to the action, and argue that such terms are not likely to help matters.
Rather than proceed to a covariant or Hamiltonian analysis of the
model in a curved spacetime, in Section~\ref{sect:plane symmetry}
we consider a plane-symmetric spacetime, and reproduce the same
propagation velocities as were observed in a flat spacetime.
We conclude in Section~\ref{sect:conclusions} with some overall comments.

\section{Flat Spacetime}
\label{sect:flat}

We begin by considering the model~\eqref{eq:genaction} in flat spacetime, since it is vastly simpler to illustrate some important dynamical aspects of the model without dealing with gravitational interactions.
That we can do so and obtain results that are applicable to curved spacetimes is not a trivial issue though, as we shall describe more fully in
Section~\ref{sect:others}.  In any case, this sort of flat
spacetime approximation (that is, so that one can transform
$\metric_{\mu\nu}\approx\eta_{\mu\nu}$ in a region of spacetime
and consider small perturbations about this background) should be
reasonable if $\metric_{\mu\nu}$ is in some sense the physically
meaningful metric of spacetime.

The Lagrange multiplier field $\lambda$ enforces the constraint:
\begin{equation}\label{eq:normalization}
A^2=A_0^2-\vec{A}^2=l,
\end{equation}
where the sign of $l$ indicates whether the vector field $A_\mu$
describes a timelike ($l>0$), null ($l=0$) or spacelike ($l<0$)
ether.  Here we are writing the spacetime covector field in terms
of a scalar and spatial vector as: $A_\mu=(A_0,\vec{A})$, and will
use, for example $\vec{A}\cdot\vec{B}=\sum_iA_iB_i$ and
$\vec{\partial}^2=\sum_i\partial_i^2$.  Variation
of~\eqref{eq:genaction} with respect to $A_\mu$ leads to the (nonlinear)
field equations:
\begin{subequations}
\begin{equation}\label{eq:vector FEQ}
\partial_\nu F^{\mu\nu}-\lambda A^\mu=J^\mu.
\end{equation}
Contraction of this with respect to $A^\mu$ and
using~\eqref{eq:normalization} yields
\begin{equation}\label{eq:l condition}
l\lambda = A_\mu\partial_\nu F^{\mu\nu}-A_\mu J^\mu,
\end{equation}
which determines $\lambda$ provided that $l\neq 0$.
Taking the divergence of~\eqref{eq:vector FEQ} yields the
constraint
\begin{equation}\label{eq:divergence}
\partial_\mu[\lambda A^\mu]=-\partial_\mu J^\mu.
\end{equation}
\end{subequations}

The first interesting feature of these models is that they possess
multiple light cones, or multiple characteristics, indicating that
there are fields in the theory that propagate at different speeds.
Since this is not obvious by cursory examination
of~\eqref{eq:vector FEQ}, we will explicitly determine these propagation velocities.

\subsection{The Reduced Timelike and Null Field Equations}

In both of these cases ($l>0$ and $l=0$ respectively) we can use~\eqref{eq:normalization} to
determine $A_0$ since the only way that $A_0$ can vanish is if
$A_\mu$ vanishes identically.

In the timelike case~\eqref{eq:l condition} will determine
$\lambda$, and the three remaining equations in~\eqref{eq:vector
FEQ} should determine $\vec{A}$.  To examine these equations we
introduce a vector $s_\mu$ such that $s_\mu A^\mu=0$, which is a
four vector $s_\mu=(s_0,\vec{s})$ with arbitrary spatial
components $s_i$, and timelike component determined from
\begin{equation}\label{eq:s0}
s_0=\vec{n}\cdot\vec{s},
\end{equation}
where we use the definition
\begin{equation}
\vec{n}=\frac{\vec{A}}{A_0}.
\end{equation}
Summing~\eqref{eq:vector FEQ} into $s_\mu$ and using~\eqref{eq:s0}
to write what results in terms of $s_i$ gives
\begin{equation}\label{eq:basic feq}
s_\mu\partial_\nu F^{\mu\nu} =s_i\bigl[ n_i\partial_\alpha
F^{0\alpha} +\partial_\alpha F^{i\alpha} \bigr] =s_i(n^iJ^0+ J^i).
\end{equation}
Since $s_i$ is arbitrary, the vector in square brackets
is the three field equations that we are seeking:
\begin{equation}\label{eq:basic feq2}
 n_i\partial_j F_{0j} +\partial_t F_{i0} -\partial_j F_{ij}
 =-(n^iJ^0+ J^i).
\end{equation}
The same equation is arrived at in the null case by solving the
zeroth component of~\eqref{eq:vector FEQ} for $\lambda$ (this is a
true algebraic equation for $\lambda$ since it involves no
second-order time derivatives of any field) and substituting it
into the spacelike equations.

Re-written in terms of the vector potential,~\eqref{eq:basic feq2}
becomes
\begin{equation}\label{eq:basic reduced}
\partial_t^2A_i
-\partial_t\partial_iA_0
-n_i\partial_t \partial_jA_j
-\vec{\partial}^2A_i
+n_i\vec{\partial}^2A_0
+\partial_i\partial_jA_j
=n_iJ^0+ J^i.
\end{equation}
These equations can be written
solely in terms of $\vec{A}$ using the derivative relations that
follow from~\eqref{eq:normalization}:
\begin{equation}
\partial_tA_0=\vec{n}\cdot\partial_t\vec{A},\quad
\partial_iA_0=\vec{n}\cdot\partial_i\vec{A},
\end{equation}
which leads to a second-order, quasi-linear partial differential equation of the form:
\begin{subequations}\label{eq:quasi linear}
\begin{equation}
L_{ij}(A_k,\partial_t^2,\partial_t\partial_k,\partial_k\partial_l)A_j
+b_i(A_k,\partial_tA_k,\partial_lA_k)=0,
\end{equation}
where
\begin{align}
L_{ij}
&=\bigl[
\delta_{ij}\partial_t^2
-n_j\partial_t\partial_i
-n_i\partial_t\partial_j
-\delta_{ij}\vec{\partial}^2
+n_in_j\vec{\partial}^2
+\partial_i\partial_j
\bigr],\\
\label{eq:bi}
b_i&=A_0^{-1}n_i(\delta_{kl}-n_kn_l)\partial_jA_k\partial_jA_l
-A_0^{-1}(\delta_{jk}-n_jn_k)\partial_iA_j\partial_tA_k.
\end{align}
\end{subequations}
While the lower-order derivative terms~\eqref{eq:bi} are important to determine the particular form of the solutions of~\eqref{eq:quasi linear}, in this case they are irrelevant when determining the characteristics (see, for example, the discussion in~\cite{Renardy+Rogers:1992}).
For this second-order, quasilinear system, the characteristic polynomial (see in particular the discussion in Section~VI.6 in~\cite{Choquet-Bruhat+:1989}) may be found by making the replacement $\partial_i\rightarrow ip_i$ and $\partial_t\rightarrow ip_0$ in $L_{ij}$, and computing the determinant
\begin{equation}
\det (L_{ij})=(p_0^2-\vec{p}^2)^2(p_0-\vec{n}\cdot\vec{p})^2=0.
\end{equation}
(This is most easily accomplished performing a rotation so that
$\vec{n}=(n_1,0,0)$ which we can do since we are working pointwise--the results can then be re-assembled to give the general
result.) 
This shows us that the initial value problem is well-posed provided we are imposing initial data on non-characteristic surfaces, where is this case the condition for a hypersurface to be non-characteristic depends on the vector field $A_i$.
This is not particularly unusual, as exactly this situation also occurs in general relativity where the condition that data be imposed on a spacelike hypersurface involves the metric tensor.
While~\eqref{eq:quasi linear} is not diagonal, and so the discussions in, for example~\cite{Wald:1984,Hawking+Ellis:1973}, do not apply directly, it may be made so following the procedure given in~\cite{Choquet-Bruhat+:1989}.

We see therefore that there are characteristics defined
by
\begin{subequations}\label{eq:flat cones}
\begin{equation}\label{eq:flat cones noncov}
p_0=\pm \lvert\vec{p}\rvert,\quad\text{and}\quad
p_0=\vec{n}\cdot\vec{p},
\end{equation}
two of which are the usual null propagation conditions, and the
third depends on nonlinear contributions from
$\vec{A}$. We therefore see quite clearly why this mode was not so
well understood when linearizing about a flat background with
$\vec{A}=0$~\cite{Jacobson+Mattingly:2000}: in the linearized
limit we would have a $p_0=0$ `characteristic', which could be
interpreted as a non-propagating field (that this mode should
propagate was recognized in~\cite{Jacobson+Mattingly:2000}).

Before considering the spacelike case, we note that we can write~\eqref{eq:flat cones noncov} in the covariant form:
\begin{equation}
p^2=0,\quad\text{and}\quad
A_\mu p^\mu=0,
\end{equation}
\end{subequations}
and so it is not too surprising that we will find exactly the same
result.
As we will see in Section~\ref{sect:shocks}, and later on in a curved spacetime in Section~\ref{sect:plane symmetry}, the field equations associated with the new characteristic surfaces have a tendency to form shocks.
Note that although we have performed this analysis in
three spatial dimensions, the result should hold in any number of
spacetime dimensions.

\subsection{The Reduced Spacelike Field Equations}

It requires a bit more effort to show that~\eqref{eq:flat cones}
also holds in the spacelike case. To begin, we write
$A_\mu=(A_0,A\hat{a})$ where $\hat{a}$ is a unit vector in the
direction of $\vec{A}$ and $A$ is the magnitude of the spatial
component (so $A=\sqrt{\vec{A}^2}$ and $\hat{a}=\vec{A}/A$).  The
normalization condition is $A_0^2-A^2=l<0$, which can be solved
for $A$ (this requires that $A>0$, and it is now possible that
$A_0=0$).

We again introduce a vector $s_\mu$ satisfying $s_\mu A^\mu=0$,
but in this case we have $s_0A_0=A\vec{s}\cdot\hat{a}$, and
therefore the vector $\vec{s}$ must have the form
\begin{equation}
\vec{s}=s_{A}\hat{a}+\vec{s}_T, \quad\text{where}\quad
\vec{s}_T\cdot\hat{a}=0.
\end{equation}
Again summing $s_\mu$ into~\eqref{eq:vector FEQ} gives only three
equations, one proportional to $s_0$ and the other two
proportional to $\vec{s}_T$:
\begin{equation}
s_0\Bigl(\partial_\nu F^{0\nu} +\frac{A_0}{A}a_i\partial_\nu
F^{i\nu}\Bigr) +s_{Ti}\partial_\nu F^{i\nu}
 =s_0\Bigl(J^0 +\frac{A_0}{A}a_iJ^i\Bigr)
+s_{Ti} J^i,
\end{equation}
which can be written as
\begin{subequations}\label{eq:lt proj}
\begin{gather}
\partial_\nu F^{0\nu} +\frac{A_0}{A}a_j\partial_\nu F^{j\nu}
 =J^0 +\frac{A_0}{A}a_jJ^j, \\
\label{eq:proj spacelike}
(\delta_{ij}-a_ia_j)\partial_\nu F^{j\nu}
 = (\delta_{ij}-a_ia_j)J^j.
\end{gather}
\end{subequations}
Multiplying the second by $A_0/A$ and adding the first multiplied
by $a_i$, we have
\begin{equation}\label{eq:spacelike reduced}
 a_i\partial_\nu F^{0\nu}
 +\frac{A_0}{A}\partial_\nu F^{i\nu}
 =a_iJ^0
 +\frac{A_0}{A}J^i,
\end{equation}
which are three equations that correctly reproduce~\eqref{eq:lt
proj}.

Since~\eqref{eq:spacelike reduced} is equivalent
to~\eqref{eq:basic feq2} up to a multiplicative factor $A_0/A$,
the analysis that we performed to find the characteristic surfaces
will proceed as before, and lead to~\eqref{eq:flat cones}.  The
analysis has properly taken care of the fact that here $\vec{A}$
cannot vanish and $A_0$ can, and we see from~\eqref{eq:flat cones}
that in the $A_0\rightarrow 0$ limit the speed of propagation of
one of the modes goes to infinity.

\section{Perturbative Analysis}
\label{sect:linearization}

We consider a generic background $A_\mu$ and perform a
linearized expansion: $A_\mu\rightarrow A_\mu=\delta A_\mu$,
$\lambda=\lambda+\delta\lambda$ and $J^\mu\rightarrow J^\mu+\delta
J^\mu$. From the expansion of~\eqref{eq:normalization} we see that
\begin{subequations}\label{eq:basic linearization}
\begin{equation}\label{eq:delta A}
A^\mu\delta A_\mu=0,
\end{equation}
regardless of the sign of $l$.  Then \textit{provided that
$A_0\neq 0$} we can proceed in a similar manner as in
Section~\ref{sect:flat}, which leads to the perturbation equations
\begin{equation}\label{eq:delta others}
 \partial_j \delta F_{ij}
 -\partial_t \delta F_{i0}
 -n_i\partial_j \delta F_{0j}
 +\lambda(\delta_{ij}-n_in_j)\delta A_j
 =\delta J^i+n_i\delta J^0.
\end{equation}
\end{subequations}
From this we see that this has a principle part that is equivalent to~\eqref{eq:basic feq2}, and
therefore perturbations will have characteristics determined
by~\eqref{eq:flat cones}.
This result is redundant given that we had already determined the characteristics in the previous section, but may be looked upon as an alternative derivation.

That~\eqref{eq:delta others} is correct unless $A_0=0$, should be a signal that there is something special about such configurations.
What we are going to show is that performing a linearization about such background solutions does not produce linearized equations that approximate exact solutions of the field equations.
This is a Cauchy instability~\cite{Hawking+Ellis:1973} in the exact theory, which shows up as a linearization instability in the perturbative calculation (see, for example~\cite{Brill+Deser:1973,Clayton:1996}).
In this section we will show that the linearized equations only have $p^2=0$ characteristics, contradicting the exact results~\eqref{eq:flat cones}.
We will also show in Sections~\ref{sect:shocks},~\ref{sect:Hamiltonian} and~\ref{sect:plane symmetry} that the field equations are singular in the $A_0\rightarrow 0$ limit.

We must first consider background solutions with $A_0=0$.  
Note that in order to write the field
equations in the form~\eqref{eq:spacelike reduced} we multiplied~\eqref{eq:proj spacelike} by $A_0$, and therefore taking $A_0=0$ in~\eqref{eq:spacelike reduced} will not result in a full set of field equations.
From~\eqref{eq:normalization} we must have $\vec{A}^2=-l>0$, so
we see that we only have two degrees of freedom remaining.
The zeroth component of~\eqref{eq:vector FEQ} results in
\begin{equation}\label{eq:long}
\partial_t\partial_i A_i=-J^0,
\end{equation}
and~\eqref{eq:l condition} determines $\lambda$ as:
\begin{equation}
l\lambda
 =A_i(\partial_i\partial_jA_j
 -\vec{\partial}^2 A_i
 +\partial_t^2 A_i)
 -A_iJ^i.
\end{equation}
Since this determines $\lambda$ we can use this to write the
remaining two spatial components of~\eqref{eq:vector FEQ} as
\begin{subequations}\label{eq:trans}
\begin{equation}
 P_{ij}(\partial_t^2A_j
 +\partial_j\partial_kA_k-\vec{\partial}^2A_j)=P_{ij}J^j,
\end{equation}
where the projection tensor is defined as
\begin{equation}
 P_{ij}=\delta_{ij}+l^{-1}A_iA_j.
\end{equation}
\end{subequations}
Note that we have three equations for the two remaining
functions in $A_i$.  
This is not surprising since we have imposed $A_0=0$ by hand, and we expect that this will only be possible with some restriction on the source $J^\mu$.
Rather than determining the general form of these restrictions, we will choose a particular solution with: $J^0=0$ and $J^i=0$, which implies that $\lambda=0$ and allows a constant $A_i$.

Performing the linearized expansion about this solution, from~\eqref{eq:delta A} we find that $A_i\delta A_i=0$.
The timelike component of the linearization of~\eqref{eq:vector FEQ} leads to
\begin{subequations}\label{eq:A0pert}
\begin{equation}
 \partial_i^2\delta A_0 
 -\partial_t\partial_i\delta A_i
 =\delta J^0,
\end{equation}
and summing $A_i$ into the linearization of the spacelike components of~\eqref{eq:delta A} yields an equation for $\delta \lambda$:
\begin{equation}
l\delta\lambda =-A_i
 \bigl(\partial_t \delta F_{i0}-\partial_j\delta F_{ij}\bigr)
 -A_i\delta J^i.
\end{equation}
Using this to remove $\delta\lambda$ from the remaining perturbation equations results in
\begin{equation}
P_{ij}
 \bigl(\partial_t \delta F_{i0}-\partial_j\delta F_{ij}\bigr)
  =-P_{ij}\delta J^j.
\end{equation}
\end{subequations}

Assuming that $A_i$ is a constant vector pointing along the $x$-direction: $A_i=(a,0,0)$
where $a^2=-l$, then~\eqref{eq:delta A}
implies that $\delta A_x=0$.  It is then a fairly straightforward
matter to re-write the field equations that result from~\eqref{eq:A0pert}
as:
\begin{subequations}
\begin{align}
 (\partial_t^2-\vec{\partial}^2)\partial_x^2\delta A_0
 &=(\partial_t^2-\partial_x^2)\delta J^0
 +\partial_t(\partial_y\delta J^y+\partial_z\delta J^z),\\
 (\partial_t^2-\vec{\partial}^2)\partial_x^2(\partial_y\delta A_y+\partial_z\delta A_z)
 &=\partial_x^2(\partial_y\delta J^y+\partial_z\delta J^z)
  +\partial_t(\partial_y^2+\partial_z^2)\delta J^0,\\
 (\partial_t^2-\vec{\partial}^2)(\partial_y\delta A_z-\partial_z\delta A_y)
 &=\partial_z\delta J^y-\partial_y\delta J^z.
\end{align}
\end{subequations}
This system contains three modes that all propagate at the speed of light,
with no sign of the mode that propagates along the $p_\mu A^\mu=0$ characteristic.

These linearized equations would therefore lead one
to believe that small fluctuations about a field configuration
with $A_0=0$ \textit{all} propagate at the speed of light.  We
know from the exact field equations that this is most definitely
\textit{not} the case, and that there is a degree of freedom that
has an infinitely large propagation velocity in the limit that $A_0=0$.
It should be reasonably clear
that this is happening because we are attempting to perform a
perturbation analysis about a degenerate field configuration--in
fact our initial data surface is a characteristic surface for the
exact field equations.

This suggests us that the expansion performed
in~\cite{Kostelecky+Samuel:1989} is suspect.  
It is, of course,
possible that other models will result in the linearized equations
that they use, but it is clear from what we have seen here that the linearized equations that they derive do \textit{not} approximate solutions to the exact field equations.
Lest one think that this sort of thing would not happen when
the model is coupled to gravity, in Section~\ref{sect:plane symmetry} we
show that the same characteristics~\eqref{eq:flat cones} appear
in a curved spacetime.

We also see that this does not happen in the null and timelike
cases, but this does not mean that there are not other issues to deal with.
We will show in the next section that shock formation should be expected, and
in Section~\ref{sect:Hamiltonian}, that the (classical) Hamiltonian of these
models will not be bounded from below.

\section{Shock Formation}
\label{sect:shocks}

Now we wish to examine some properties of nontrivial solutions that will display propagation along the $p_\mu A^\mu=0$ characteristic.
One such example is that of a plane-symmetric vector potential:
$A_\mu=(A_0(t,x),A_1(t,x),0,0)$ (and correspondingly for the source), since
these solutions are expected to propagate along the $x$-direction, along which $\vec{A}$ also points.
What we will discover (for a non-null ether) is that the field equations reduce
to a form that can be written as a conservation law and is generically
expected to form shocks.

From~\eqref{eq:vector FEQ} we find the two field equations:
\begin{subequations}
\begin{align}
 \partial_x^2A_0-\partial_t\partial_x A_1
 -\lambda A_0&=J^0,\\
 \partial_t^2A_1-\partial_t\partial_x A_0
 +\lambda A_1&=J^1,
\end{align}
\end{subequations}
and taking a linear combination that removes $\lambda$ we find:
\begin{equation}
 A_1\bigl(\partial_x^2A_0-\partial_t\partial_x A_1\bigr)
+A_0\bigl(\partial_t^2A_1-\partial_t\partial_x A_0\bigr)
 =A_1J^0+A_0J^1.
\end{equation}
Using~\eqref{eq:normalization}, this can be written as an equation determining  either $A_0$ or $A_1$, both forms of which are:
\begin{subequations}\label{eq:flat ps eqns}
\begin{align}
\label{eq:flat ps A1}
 \Bigl(\partial_t-\frac{A_1}{A_0}\partial_x\Bigr)
 \Bigl(\partial_tA_1-\frac{A_1}{A_0}\partial_xA_1\Bigr)&=\frac{A_1}{A_0}J^0+J^1,\\
\label{eq:flat ps A0}
 \Bigl(\frac{A_0}{A_1}\partial_t-\partial_x\Bigr)
 \Bigl(\frac{A_0}{A_1}\partial_tA_0-\partial_xA_0\Bigr)&=J^0+\frac{A_0}{A_1}J^1.
\end{align}
\end{subequations}
In either case we see that in a region external to the sources we need only consider a homogeneous, first-order, partial differential equation, solutions to which are also solutions when the source is non-vanishing.

For a null ether we have $A_0=\pm A_1$, and~\eqref{eq:flat ps A1} leads to
\begin{subequations}\label{eq:shock equations}
\begin{equation}\label{eq:null shocks}
\partial_tA_1\mp \partial_x A_1=0,
\end{equation}
with solutions: $A_1=A_1(t\pm x)$.
Throughout we will assume that $A_0>0$
(the ether is forward-pointing in time), and so disturbances in the ether
field propagate in the opposite spatial direction to which the ether field itself points.
This last fact is a generic feature, consistent with the characteristic surface defined by $p_\mu A^\mu=0$ in~\eqref{eq:flat cones}.

For a timelike either we can solve~\eqref{eq:normalization} for $A_0$ as: $A_0=\pm \sqrt{A_1^2+l}$, and so we consider solutions of:
\begin{equation}\label{eq:timelike shocks}
\partial_tA_1\mp \frac{A_1}{\sqrt{A_1^2+l}}\partial_x A_1
 =\partial_tA_1\mp \partial_x \sqrt{A_1^2+l}
 =0,
\end{equation}
where the second form is written as a conservation law.
This type of equation is considered in many standard texts on
fluid or gas dynamics (see for
example~\cite{Renardy+Rogers:1992,Chorin+Marsden:1992}), from
which it is evident that, for example, a Gaussian fluctuation of $A_1$
about a constant background $A_0$ will inevitably lead to shock
formation.  The value of $A_1$ is constant along the
characteristics, which are straight lines determined from the
initial value of $A_1$ by: $x(t)=k(x_0)t+x_0$ where $k(x_0)=\mp
A_1(x_0)/\sqrt{A_1^2(x_0)+l}$ and $A_1(x_0)=A_1(0,x_0)$.
Note though that since $l>0$ we
have $\lvert k\rvert < 1$, and disturbances propagate at a
velocity slower than that of light.  

For a spacelike either we have a similar situation.  Solving~\eqref{eq:normalization} for $A_1$ and using~\eqref{eq:flat ps A0} to determine $A_0$, we find (this can also be written as a conservation law):
\begin{equation}\label{eq:spacelike shocks}
\partial_tA_0\mp \frac{\sqrt{A_0^2-l}}{A_0}\partial_x A_0
 =0,
\end{equation}
\end{subequations}
which shows that disturbances not only
generically form shocks but that they also travel faster than
light.
We also see quite clearly that the limit $A_0\rightarrow 0$ does not lead to a smooth linear limit.
Writing the characteristics as above, we now have $k(x_0)=\mp \sqrt{A_0^2-l}/A_0$, and regions with very small $A_0$ have characteristics that approach infinite velocity.
Since disturbances in $A_0$ cannot be assumed to be exactly constant over an extremely large region, shock formation is to be expected.

Although we have explored the system in a plane-symmetric
situation (which we shall explore further in
Section~\ref{sect:plane symmetry} in a curved spacetime),
we have found that an identical result holds for
spherically-symmetric solutions.  Although a spherically-symmetric
ether is difficult to motivate, it does show that the existence of
shocks is not an artifact of the specific solution
that we are considering.

At this point it is worthwhile saying a few words about a
restricted gauge invariance that the system possesses when the external source coupling is conserved (that is, when $\partial_\mu J^\mu=0$).
In this case the
action~\eqref{eq:genaction} is invariant under the gauge
transformation $A_\mu\rightarrow A_\mu+\partial_\mu\theta$
provided that $\theta$ satisfies:
$\metric^{\mu\nu}(A_\mu+\partial_\mu\theta)(A_\nu+\partial_\nu\theta)=l$.
In the present model, this simplifies to (solving for
$\partial_t\theta$):
\begin{equation}
\partial_t\theta=-A_0\pm\sqrt{A_0^2+\partial_x\theta(\partial_x\theta+2A_1)}.
\end{equation}
It can be shown that all of~\eqref{eq:shock equations} are invariant under this
transformation, so that $A_\mu$ may be transformed to zero by an
appropriate choice of gauge.

This should not be too surprising, since in this case the
model is identical to a Maxwell field in the nonlinear gauge: $A^2=l$.
Therefore, if the ether field is only coupled to a conserved current then the model is
gauge-equivalent to a Maxwell field, and there would be little point
working in the nonlinear gauge~\eqref{eq:normalization} when some more
well-behaved linear gauge (such as the Lorenz gauge $\partial_\mu A^\mu=0$)
would equivalent and far be simpler to work with.
The intended interpretation of the model is that $A_\mu$ should
represent an ether field, and therefore it will couple to matter fields
as a timelike metric representing the preferred frame: $\metric^{\mathrm{pf}}_{\mu\nu}\sim A_\mu A_\nu$~\cite{Jacobson+Mattingly:2000,Jacobson+Mattingly:2000a}, and so coupling to a conserved current is neither expected nor desired.
For the ether to have observable consequences it must couple to a non-conserved current, in which case the $p_\mu A^\mu=0$ characteristic is physical, and shock formation is expected.

\section{Canonical Analysis}
\label{sect:Hamiltonian}

The existence of multiple light cones and shock formation in a
theory of the ether is perhaps not sufficient reason
to discard the model.  The theories that we are
considering here have another dangerous feature though: they do not have a classical Hamiltonian that is bounded from below in flat
spacetime.  This analysis should be enough to convince the reader
that in any (stable) region of spacetime that can be approximated
by a flat spacetime, these theories should be generically unstable
once quantum or other dissipative effects are included.

In a quantum scenario, the standard argument is that quantum fluctuations can result in the creation of real (as opposed to virtual), negative energy particles, since no energy need be `borrowed' from the vacuum to create them. 
Thus the quantum system will cascade into states of increasingly more negative energy.
While it is possible to claim that these negative energy states are not realized in a classical model (if the initial data has finite energy it will remain with the same finite energy), vacuum fluctuations in the quantum theory seem unavoidable.

The analysis is fairly standard, so we will be brief.  
Expanding the Lagrangian
from~\eqref{eq:genaction} we find
\begin{equation}
L=\int d^3x\,\Bigl[
 \frac{1}{2}(\partial_tA_i-\partial_i A_0)^2
 -\frac{1}{4}(F_{ij})^2
 +\frac{1}{2}\lambda(A_0^2-\vec{A}^2-l)
 +A_0J^0+\vec{A}\cdot\vec{J}
\Bigr],
\end{equation}
and we see that we can treat $A_0$ and $\lambda$ as Lagrange
multipliers, and the conjugate momentum for $\vec{A}$ is
\begin{equation}\label{eq:conjugate}
P_i=\frac{\delta L}{\delta \partial_t
A_i}=\partial_tA_i-\partial_i A_0.
\end{equation}
The Hamiltonian is then
\begin{equation}\label{eq:flat Ham}
\begin{split}
H&=\int d^3x P_i\partial_t A_i-L\\
&=\int d^3x\,\Bigl[
 \frac{1}{2}\vec{P}^2
 +\vec{P}\cdot\vec{\nabla}A_0
 +\frac{1}{4}(F_{ij})^2
 -\frac{1}{2}\lambda (A_0^2-\vec{A}^2-l)
 -A_0J^0-\vec{A}\cdot\vec{J}
\Bigr],
\end{split}
\end{equation}
and we will assume that the fields fall off fast enough as
$r\rightarrow\infty$ that no surface contributions are required.

The condition imposed by $\lambda$ is~\eqref{eq:normalization},
which can be used to determine the Lagrange multiplier $A_0$ as
\begin{equation}\label{eq:A0 constraint}
A_0\approx \pm\sqrt{\vec{A}^2+l}.
\end{equation}
This is a delicate point since, as noted before, if $l<0$ there is no guarantee that $A_0\neq 0$, but since~\eqref{eq:normalization} involves the the Lagrange multiplier $A_0$ it cannot be considered as a constraint.
The condition imposed by $A_0$ is
\begin{equation}\label{eq:pseudo Gauss}
\frac{\delta H}{\delta A_0}=
 -\vec{\nabla}\cdot \vec{P} -\lambda A_0
 -J^0 \approx 0,
\end{equation}
so that since $A_0$ is already determined, this determines the Lagrange multiplier $\lambda$ as:
\begin{equation}\label{eq:lambda constraint}
\lambda \approx \mp \frac{\vec{\nabla}\cdot
\vec{P}+J^0}{\sqrt{\vec{A}^2+l}}.
\end{equation}

Evaluating the Hamiltonian~\eqref{eq:flat Ham}
``on-shell'' (replacing the Lagrange
multipliers by their phase space representation using~\eqref{eq:A0
constraint} and~\eqref{eq:lambda constraint}) in a region where
the source terms vanish, we find the remaining contributions:
\begin{equation}
H \approx\int d^3x\,\Bigl[
 \frac{1}{2}\vec{P}^2
 \pm\frac{P_iA_j\nabla_iA_j}{\sqrt{\vec{A}^2+l}}
 +\frac{1}{4}(F_{ij})^2
\Bigr],
\end{equation}
where we have used $\partial_i A_0=\pm(A_j\partial_i
A_j)/\sqrt{\vec{A}^2+l}$.  To show that this can be arbitrarily
negative on phase space, consider configurations derivable from a
static scalar potential as $A_i=\partial_i\phi(\vec{x})$.  The
conjugate momentum to $\vec{A}$ is $P_i=-\partial_iA_0$, and evaluating
the Hamiltonian density yields: $-(1/2)(\partial_iA_0)^2$, which
can be made arbitrarily large and negative by choosing $\phi$ to
be a Gaussian with large amplitude.  Note how this has occurred:
the imposition of~\eqref{eq:normalization} has caused~\eqref{eq:pseudo Gauss},
which would have been the Gauss constraint in Maxwell's theory and led to a
positive-definite term in the Hamiltonian, to now become an equation that
determines a Lagrange multiplier, and the Hamiltonian is indefinite.

We can also consider Hamilton's equations, which appear fairly innocent
when written in terms of the Lagrange multipliers:
\begin{subequations}\label{eq:flat Hams}
\begin{align}
\dot{A}_i=\{A_i,H\}=\frac{\delta H}{\delta P_i}
 &=P_i+\partial_iA_0
 \approx P_i
 \pm \frac{A_j\partial_i A_j}{\sqrt{\vec{A}^2+l}}, \\
\label{eq:flat Hams P}
\dot{P}_i=\{P_i,H\}=-\frac{\delta H}{\delta A_i}
 &= -\partial_j F_{ij}
 -\lambda A_i+J^i\nonumber \\
 &\approx -\partial_j F_{ij}
 \pm A_i\frac{\vec{\nabla}\cdot
\vec{P}+J^0}{\sqrt{\vec{A}^2+l}}
 +J^i,
\end{align}
\end{subequations}
where in the second form of these~\eqref{eq:A0 constraint}
and~\eqref{eq:lambda constraint} have been imposed. 
Note that as far as the initial
value formulation is concerned the sign of $l$ is largely
irrelevant, but manifests itself by allowing the possibility
that $A_0=0$ in the spacelike case, in which limit~\eqref{eq:flat Hams P} is singular. 
This is another reflection
of the fact that in this model $A_0=0$ is a special point, and we
cannot expect that it may be reached smoothly (see
also~\cite{Isenberg+Nester:1977}).

\section{Related Models}
\label{sect:others}

Before demonstrating that all of the aforementioned issues also
arise in the curved spacetime form of the model, we make a few comments on
flat spacetime generalization of~\eqref{eq:genaction}.

If we were to introduce non-Maxwell kinetic contributions such as $(\nabla_\mu A^\mu)^2$ into~\eqref{eq:genaction} as suggested in~\cite{Jacobson+Mattingly:2000a}, we would perhaps have
different characteristics, but all of the difficulties that we have been discussing here would be present. 
Since the model will have time derivatives of $A_0$ in the action, $A_0$
is no longer a Lagrange multiplier and the
$\vec{P}\cdot\vec{\partial}A_0$ term remains in the Hamiltonian,
along with new contributions
$P_0^2/(2d)+P_0\vec{\partial}\cdot\vec{A}$ involving $P_0$ (the momentum
conjugate to $A_0$).  In this case since both $P_0$ and $A_0$ are
phase space coordinates, and it is a trivial matter to find field
configurations on which the Hamiltonian is as large and negative
as one likes.  In order for this theory to be equivalent to a
Maxwell field written in a nonlinear gauge one would need a
restricted gauge invariance that simultaneously left the
$(\nabla_\mu A^\mu)^2$ term and the
condition~\eqref{eq:normalization} invariant, which seems unlikely.  
That this type of model is
`dangerous' in this sense has already been discussed in the
literature~\cite{Isenberg+Nester:1977}.
The same reference also shows that the model considered as an example of a preferred-frame theory of gravity in~\cite{Will+Nordtvedt:1972a}, has stability issues that go beyond the PPN analysis performed there.

Similar comments apply to the possible addition of
curvature-coupling terms such as $R_{\mu\nu}A^\mu A^\nu$. 
Since this would constitute an explicit
derivative-coupling, altered characteristics are to be
expected~\cite{Kuchar:1977}, and were observed
in~\cite{Will:1993}. 
In the latter reference the
characteristic surfaces of the exact theory are not determined, and so
it is not known whether the perturbation that is performed is
degenerate in the sense discussed in
Section~\ref{sect:linearization}.

If we alter the term that imposes the constraint~\eqref{eq:normalization} to $\lambda(A^2-l)^2$ as suggested in~\cite{Kostelecky+Samuel:1989}, then $\lambda$ will drop out of the field equations altogether,
leaving~\eqref{eq:normalization} and $\partial_\nu
F^{\mu\nu}=J^\mu$. Consistency of the field equations \textit{requires} that the vector source is conserved: $\partial_\mu J^\mu=0$, and since the
divergence equation is trivial there are really only four field
equations. Since the coupling is therefore required to be gauge-invariant, this is a
Maxwell-theory written in a nonlinear gauge defined
by~\eqref{eq:normalization}. 
Therefore the $n=2$ model of~\cite{Kostelecky+Samuel:1989} is just Einstein-Maxwell theory in disguise, with the Lorentz symmetry breaking as the result of a nonlinear gauge choice for the vector field.
It is interesting that adding a
self-coupling potential term of the form $V(A^2)$ to this theory could have disastrous
consequences, since even if the coupling current is conserved the
divergence of the field equations lead to $\partial_\mu A^\mu=0$,
which is a nontrivial equation determining a degree of freedom in
$A_\mu$, therefore there are five field equations for
the four unknown functions $A_\mu$.  In contrast, introducing a
self-interaction term to~\eqref{eq:genaction} merely induces the
shift: $\lambda\rightarrow \lambda + V^\prime(l)$. This
demonstrates that the inclusion (or not) of a mass term in a
vector field theory can have rather significant consequences,
contrary to the statement made in~\cite[Section~$5.4$]{Will:1993}.

We should mention that the vector field-based model that we have introduced~\cite{Clayton+Moffat:1999} is of a different class than those discussed here.  
Because we are imposing no additional constraints in addition to those already present in an Einstein-Proca model, we do not expect the issues discussed in this paper to arise.
On the other hand, since the vector field is not guaranteed to be timelike in these models (although this turns out to be the case in a cosmological setting), it is not so obvious that we can interpret the vector field as representing a dynamical ether fluid in general.

\section{Plane-Symmetric Spacetimes}
\label{sect:plane symmetry}

We will now examine the model in a particular curved spacetime,
demonstrating that the characteristics~\eqref{eq:flat cones},
existence of shocks, and problems performing a linear expansion
near $A_0=0$ all survive. This is a curved spacetime
generalization of the plane-symmetric solutions considered in
Section~\ref{sect:plane symmetry}, which is general enough to
contain the static background solutions that were considered
in~\cite{Kostelecky+Samuel:1989}.  We can therefore examine some nonlinear solutions `near' this spacetime.

Working now in a curved spacetime, from~\eqref{eq:genaction} we find the field equations for $A_\mu$:
\begin{subequations}\label{eq:curved feq}
\begin{equation}
\metric^{\mu\nu}A_\mu A_\nu=l,\quad
%
%
 \nabla_\nu F^{\mu\nu} -\lambda A^\mu =0,
\end{equation}
and the Einstein equations:
\begin{equation}
 G_{\mu\nu}
 =-\frac{1}{2}\Bigl(F_{\mu\alpha}{F_\nu}^\alpha
 -\frac{1}{4}\metric_{\mu\nu}F^2\Bigr)
 +\frac{1}{2}\lambda A_\mu A_\nu.
\end{equation}
\end{subequations}

The spacetimes we are interested in have an `ether
wind' along one direction:
\begin{equation}
A_\mu=(A_0(t,x),A_1(t,x),0,0,\ldots,0),
\end{equation}
consistent with translation and rotation killing vectors in a flat
$n$-dimensional hypersurface, with plane-symmetric metric:
\begin{equation}
\metric_{\mu\nu}
 =\begin{pmatrix}
 \metric_{ab}(t,x) & 0 \\
 0 & -\alpha^{4/n}(t,x)\delta_{ij}
 \end{pmatrix},
\end{equation}
where $a,b,\ldots\in\{1,2\}$ and $i,j,\ldots \in\{3,4,\ldots,
n+2\}$, so that $n$ is the dimension of the flat subspace.  We use
coordinates $\{y^i\}$ in the flat hypersurface and $t$ and $x$ in
the `curved' directions. We will also use the remaining coordinate
invariance in the $t-x$ plane to choose the conformally-flat
metric
\begin{equation}
\metric_{ab}=e^\phi\eta_{ab},
\end{equation}
where $\eta_{ab}$ is the two-dimensional Minkowski metric: $\eta_{ab}=\mathrm{diag}(1,-1)$.

Reducing the action results in (dividing out the infinite volume
$\int d^ny$ in the flat plane):
\begin{multline}\label{eq:cflat action}
 S_{\mathrm{red}}=\int dt\,dx\,\Bigl[
 -2\alpha\eta^{ab}\partial_a\alpha\partial_b\phi
 -\frac{4(n-2)}{n}\eta^{ab}\partial_a\alpha\partial_b\alpha \\
 -\frac{1}{4}\alpha^2 e^{-\phi}F_{01}^2
 +\frac{1}{2}\lambda \alpha^2(\eta^{ab}A_aA_b-le^\phi)
 \Bigr],
\end{multline}
where
\begin{equation}
F_{01}=\partial_tA_1-\partial_xA_0.
\end{equation}
From this we find the field equations
\begin{subequations}\label{eq:plane symmetric feq}
\begin{gather}
(\partial_t^2-\partial_x^2)\alpha^2 +\alpha^2\Bigl(\frac{1}{4}e^{-\phi}F_{01}^2
-\frac{1}{2}l\lambda e^\phi\Bigr)=0,\\
(\partial_t^2-\partial_x^2)\phi
+\frac{4(n-2)}{n}\frac{1}{\alpha}(\partial_t^2-\partial_x^2)\alpha
 -\frac{1}{2}e^{-\phi}F_{01}^2=0,\\
\label{eq:ether constraint}
A_0^2-A_1^2-le^\phi=0,\\
 \partial_x\bigl[\alpha^2e^{-\phi}F_{01}\bigr]
 -2\lambda\alpha^2 A_0=0,\\
 \partial_t\bigl[\alpha^2e^{-\phi}F_{01}\bigr]
 -2\lambda\alpha^2 A_1=0,
\end{gather}
\end{subequations}
which are identical to what we would have found directly from the
field equations~\eqref{eq:curved feq}.  The linear combination of
the last two of these from which $\lambda$ cancels gives
\begin{equation}\label{eq:ps general}
 A_0\partial_t\bigl[\alpha^2e^{-\phi}F_{01}\bigr]
 -A_1\partial_x\bigl[\alpha^2e^{-\phi}F_{01}\bigr]=0,
\end{equation}
and we see that as expected, the $p_\mu A^\mu =0$ characteristic has
survived.

Although~\eqref{eq:ps general} results in $\alpha^2e^{-\phi}F_{01}=\mathit{constant}$, along these characteristics, it is
difficult to solve the remaining equations.
We will instead consider the
simpler class of solutions with $F_{01}=0$, which also implies
that $\lambda=0$. In this case we find (writing the equation in
both forms analogous to~\eqref{eq:flat ps eqns})
\begin{subequations}\label{eq:pss}
\begin{align}
A_0\partial_tA_1-A_1\partial_xA_1&=\tfrac{1}{2}le^\phi\partial_x\phi,\\
A_0\partial_tA_0-A_1\partial_xA_0&=\tfrac{1}{2}le^\phi\partial_t\phi.
\end{align}
\end{subequations}
Both of $\alpha$ and $\phi$ will be functions of either $t+x$ or
$t-x$, and for the the solutions $\phi$ that are either right- or
left-moving  ($\phi=\phi(t\mp x)$), we find (again assuming for
forward-pointing $A_0>0$) that $A_1$ propagates in the same
direction as $\phi$. The equations~\eqref{eq:pss} then become
($\phi^\prime =\partial_t\phi(t\mp x)$):
\begin{subequations}
\begin{align}
\label{eq:psts}
\partial_tA_1\mp \frac{A_1}{\sqrt{A_1^2+l}}\partial_xA_1
 &=\tfrac{1}{2}le^\phi\frac{\phi^\prime}{\sqrt{A_1^2+l}},\\
\label{eq:psss}
\partial_tA_0\mp \frac{\sqrt{A_0^2-l}}{A_0}\partial_xA_0
 &=\tfrac{1}{2}le^\phi\frac{\phi^\prime}{A_0},
\end{align}
\end{subequations}
and the gravitational wave $\phi$ is unaffected by the presence of
the ether.

We have the following scenarios for the different types of ether:
For a null ether all fields are decoupled, and all travel at the
speed of light. For a timelike ether, the gravitational wave
$\phi$ acts as a source in~\eqref{eq:psts}, disturbing the
ether. In this case disturbances in the ether travel slower than
$\phi$, and we generically expect shock formation in the ether. 
For a spacelike ether we have a
similar situation, but~\eqref{eq:psss} tells us that the passing
of the gravitational wave produces ether fluctuations that travel
\textit{faster} than light, and therefore faster than the
gravitational wave disturbance itself.
Similar to the flat spacetime case discussed in Section~\ref{sect:shocks}, the limit $A_0\rightarrow 0$ does not allow a linearization that represents the behavior of $A_0$ as determined by the exact field equation~\eqref{eq:psss}.

It would be interesting to explore more general solutions
to~\eqref{eq:plane symmetric feq} to find out how the
back-reaction affects the propagation of gravitational waves.
Although for the simple $F_{01}=0$ solutions that we have
considered it is a fairly simple matter to see that it is the
passing of the gravitational wave that is disturbing the ether, in
the more general case this is not obvious.
Since this would probably require numerical analysis and since the
speed of light becomes very large as $A_0\rightarrow 0$,
satisfying the Courant-Friedrichs-Lewy condition could be a
delicate business.

\section{Discussion and Conclusions}
\label{sect:conclusions}

Although the nonlinear constraint imposed in these models is algebraic ($A^2=l$ for some fixed $l$), we have shown that it nonetheless alters the causal evolution of the vector field. 
This is well-known when working in non-covariant gauges in Maxwell electrodynamics, where the resulting light cones should have no physical consequences.
In the case at hand, the (co-)vector field $A_\mu$ is intended to
represent a fundamental ether frame, giving a dynamically-determined, timelike, null or spacelike direction in spacetime.  
If this is the case, then it should not be observationally equivalent to a Maxwell field, the above constraint will have nontrivial consequences, and therefore the multiple light cones that we have shown exist in such models will be observable.

We have also shown that these models have a classical Hamiltonian that is not bounded from below.  
Although not necessarily a concern in classical conservative systems, we have to expect that any proposed quantization should seriously address the existence of a stable quantum vacuum.
We have argued that things become worse in this respect when terms like $(\nabla_\mu A^\mu)^2$ are introduced into the action, since then all four
components of $A_\mu$ would have conjugate momenta, and we have even more freedom to find classical field configurations of arbitrarily large, negative energy.
In addition, given the analysis of propagation speeds in linearized analyses of similar models~\cite{Will:1993}, a simple light cone structure is not to be expected.
Introducing curvature-coupling terms into the action cannot be expected to alleviate matters since the derivative-coupling that results is known to introduce altered propagation speeds for different fields~\cite{Kuchar:1977}, which has been explored in very similar vector field models in~\cite{Isenberg+Nester:1977}.

One of the main messages of this work is that it is important to realize that a linearized analysis does \textit{not} necessarily tell the whole story (\textit{i.e.}, a theory may have a `linearization instability'). 
In the case of a spacelike ether we saw that it is quite possible to formally expand about a degenerate configuration, and although we know from a more complete analysis that there is a field with an infinite propagation velocity, this never appears in the linearized analysis.
The canonical analysis has also shown that the Hamilton equations became singular at these solutions, and we therefore cannot expect that the resulting linearized equations to bear any resemblance to solutions of the exact field equations.
This is precisely what was noticed in earlier work on vector fields coupled to gravity~\cite{Isenberg+Nester:1977}, and shows that the analysis in~\cite{Kostelecky+Samuel:1989} is suspect.
The same concerns may arise when considering the linearized analysis of vector models in~\cite{Will:1993}, since the models are derivatively-coupled and the characteristics have not been determined.

In any case, the quantization of any of the models that we have discussed here is delicate, since the classical theory does not possess simple characteristics, fields in model are shock-forming, the classical energy is not bounded from below, and the model possesses linearization instabilities.

\section*{acknowledgments}

The author wishes to thank the Department of Physics at Acadia
University for employment during which this work was completed.


\end{document}